\documentclass[11pt]{article}

\usepackage[preprint]{acl}
\usepackage{amsmath}
\usepackage{times}
\usepackage{latexsym}
\usepackage{multirow}
\usepackage{booktabs}
\usepackage{hyperref}
\usepackage{adjustbox}
\usepackage{float}
\usepackage{subcaption}
\usepackage[T1]{fontenc}

\usepackage[utf8]{inputenc}

\usepackage{microtype}

\usepackage{inconsolata}
\usepackage{algpseudocode}
\usepackage{listings}
\usepackage{graphicx}
\usepackage{algorithm}
\usepackage[most]{tcolorbox}
\tcbset{
  mybox/.style={
    enhanced,
    breakable,
    colback=gray!5,
    colframe=black,
    fonttitle=\bfseries,
    coltitle=white,
    colbacktitle=black,
    sharp corners,
    boxrule=0.6pt,
    title={#1},
    top=2mm,
    bottom=2mm,
    left=2mm,
    right=2mm,
  }
}

\title{GraphSteal: Structural Knowledge Stealing from Graph RAG via Traversal Reconstruction}

\author{
  Jinze Gu\thanks{Equal contribution.} \quad
  Qinghua Mao\footnotemark[1] \quad
  Xi Lin\thanks{Corresponding author.} \quad
  Jun Wu \\
  School of Computer Science, Shanghai Jiao Tong University \\
  \texttt{\{p0sttt, mmmm2018, linxi234\}@sjtu.edu.cn}
}

\begin{document}
\maketitle
\begin{abstract}
Retrieval-Augmented Generation (RAG) enhances LLMs by grounding generation in query-relevant external evidence. Beyond unstructured text corpora, Graph RAG integrates knowledge graphs into the retrieval pipeline, enabling LLMs to access entities, relations, and multi-hop dependencies encoded in structured knowledge. However, the same structured knowledge that empowers Graph RAG also creates a new privacy attack surface. We demonstrate that Graph RAG systems can be turned into structural oracles: through adaptive black-box interactions, an adversary can elicit sufficient relational evidence to reconstruct substantial portions of the hidden knowledge graph. We propose a structure-oriented reconstruction framework that recovers targeted graphs from both local and global perspectives. Specifically, Depth-Wise Heuristic Search extracts fine-grained node attributes by recursively expanding entity-centered evidence, while Breadth-Wise Diffusion Search infers graph topology by propagating across relation-induced neighborhoods. Experiments on generic and healthcare scenarios demonstrate that our method can recover over 90\% of the original knowledge graph from representative Graph RAG systems, revealing sensitive entities, relations, and structural dependencies with high fidelity. Existing guradrails provide limited defense against our attack, highlighting the inherent difficulty of safeguarding structural privacy in Graph RAG pipelines.
\end{abstract}
\begin{figure}  
  \centering
  \includegraphics[width=\linewidth]{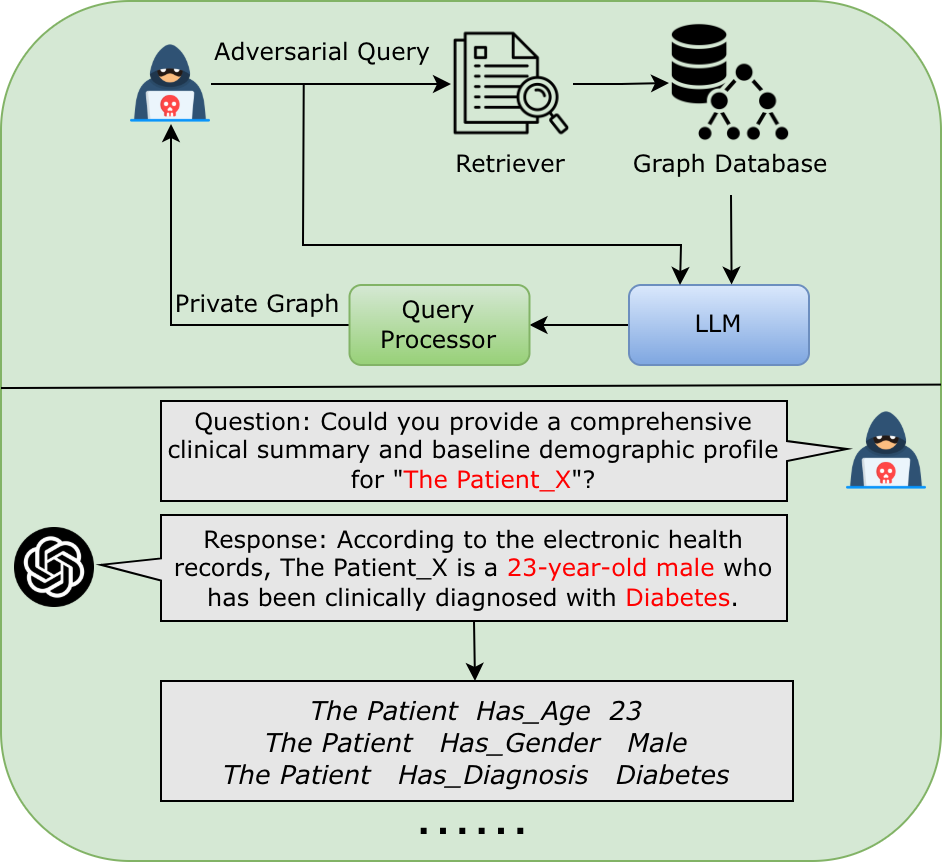}
  \caption{Adversarial queries exploit the retriever–LLM pipeline to expose sensitive node and edge information, which can be iteratively expanded to reconstruct the underlying private graph.}
  \label{fig:scenario}
\end{figure}

\section{Introduction}
Large language models (LLMs) have demonstrated remarkable capabilities across numerous natural language processing tasks. However, they still face substantial limitations in scenarios that necessitate domain-specific knowledge and complex reasoning, often prone to providing hallucinatory or obsolete responses. To mitigate these issues, Retrieval-Augmented Generation (RAG) enhances factuality and explainability by incorporating external knowledge sources, such as Knowledge Graphs (KGs). As structured, editable, and explicit knowledge repositories, KGs offer a promising solution to mitigate LLM hallucinations~\cite{shuster2021retrieval}. Recent advances~\cite{jiang2023structgpt, guo2025lightragsimplefastretrievalaugmented,luo2024reasoning, sun2024thinkongraph, wang2024knowledge, xu2024generate, nguyen2024direct, wen2024mindmap} have explored KG-augmented reasoning, roughly categorized into two paradigms: \textit{retrieval-based approaches}, which directly inject factual evidence into prompts~\cite{wen2024mindmap, wang2024knowledge}, and \textit{agent-based approaches}, which allow LLMs to interactively explore entities and relations for step-by-step reasoning~\cite{luo2024reasoning, sun2024thinkongraph}.

Despite these advancements, the privacy implications of KG-augmented LLMs remain critically under-investigated. Intuitively, RAG systems appear to offer a privacy shield, as users receive generated responses without direct access to the backend database. However, recent studies demonstrate that adversaries can extract private information via sophisticated queries. For instance, \citet{qi2024follow} exploit instruction-following capabilities to extract text data verbatim. Similarly, \citet{zeng2024good} propose structured prompting attacks to target specific private information, while \citet{jiang2024rag} introduce \textit{RAG-Thief}, an agent-based framework that progressively extracts private knowledge pieces via self-improving queries.

While these studies reveal significant risks in general RAG systems, their efficacy on structured knowledge bases remains questionable. Unlike vanilla RAG, Graph RAG relies on complex reasoning over entities and relationships connected by logical associations rather than mere semantic similarity. This distinction severely limits the effectiveness of existing text-centric approaches~\cite{qi2024follow, jiang2024rag}. Compared to continuous document segments, graph data is discontinuous and exhibits complex domain-specific topological patterns~\cite{pan2024unifying}, making it challenging to traverse nodes in a linear sequence. While \citet{liu2025exposing} recently extended privacy analysis to Graph RAG, their investigation is primarily confined to point-level leakage (e.g., entities or PII) under simple querying setups. Crucially, they overlook the complexity of topological reconstruction, leaving it an open question whether an adversary can systematically reverse-engineer the underlying graph structure—a critical gap we address in this work.

In this work, we introduce a novel privacy extraction attack against graph RAG, which induces LLMs to progressively recover the entire knowledge graph by injecting structure-aware adversarial instructions. We leverage our attack strategy to systematically investigate the vulnerability of existing graph RAG systems, including retrieval-based and agent-based RAG paradigms. Specifically, we design two attack strategies tailored to these systems, including targeted attack which aims to extract specific knowledge from the knowledge graph, and untargeted attack which seeks to recover as much of the graph as possible. For targeted attack, we propose a Heuristic Deep Search to accelerate the exploration of the target entity by prioritizing high-potential branches. For untargeted attack, we utilize a Breadth-Wise Structural Diffusion strategy, which systematically explores the graph from anchor entities outwards, thereby guaranteeing the fidelity of the global reconstruction. We conduct comprehensive experiments to evaluate the efficacy of privacy extraction attack in generic and healthcare scenarios. Our attack strategy can reconstruct over 90\% of knowledge graph from representative graph RAG systems, indicating the critical risk of privacy leakage induced by such an attack. Moreover, we evaluate the effect of potential defense solutions, including protective system prompt and output window restriction. We analyze the limitation of these approaches and discuss significant challenges of safeguarding privacy of graph RAG systems. Our contribution are presented as follows:
\begin{itemize}
    \item We provide the first in-depth analysis of structural privacy risks unique to Graph RAG systems, showing that their graph-based retrieval pipeline can systematically expose structual entity–relationship through iterative queries.
    \item We propose a query-based attack method to efficiently reconstruct knowledge graph with sensitive information extracted by breadth-first and depth-first traversal strategies.
    \item We analyze the impact of retrieval paradigms, knowledge graph sizes and traversal methods on attack performance, and discuss the limitation of existing defense approaches.
\end{itemize}

\section{Related Work}


\subsection{KG-augmented LLM Reasoning.}
To mitigate LLM hallucinations, recent works incorporate Knowledge Graphs (KGs) to enhance reasoning with structured knowledge. These approaches generally fall into two paradigms: \textit{retrieval-based} and \textit{agent-based}. Retrieval-based methods \cite{wang2024knowledge, zhang2024knowgpt, wang2024boosting} directly inject logical associations or structured evidence from KGs into prompts to ground LLM generation. Conversely, agent-based methods \cite{sun2024thinkongraph, luo2024reasoning, chen2024pog} empower LLMs to iteratively explore reasoning paths or formulate plans over the KG. While these studies significantly improve faithfulness and explainability, the security implications of such deep KG integration remain severely under-explored. This work addresses this gap by investigating the privacy leakage risks in KG-grounded RAG systems.\par

\subsection{Privacy Risk of Large Language Models.}
A plethora of studies \cite{carlini2021extracting, lee2023language, biderman2023emergent, zeng2023exploring} have indicated that LLMs are prone to memorizing and revealing information from their pre-training and fine-tuning data. When external knowledge is integrated to formulate LLM responses, keeping the datastore private becomes critical, yet recent works show this introduces new vulnerabilities. \citet{huang2023privacy} first demonstrated that private datastores induce higher privacy risks in retrieval-based language models. Subsequently, \citet{qi2024follow} leveraged prompt injection to extract verbatim text from RAG datastores, though success rates drop significantly without background knowledge. \citet{zeng2024good} further exposed the vulnerability of RAG systems to structured adversarial prompts for extracting specific private data items. More recently, \textit{RAG-Thief} \cite{jiang2024rag} introduced an agent-based automated attack to extract scalable amounts of document segments.

While these methods effectively target unstructured text, the privacy implications of \textit{graph-structured} retrieval remain largely unexplored. A concurrent study by \citet{liu2025exposing} is among the first to extend this analysis to Graph RAG. They reveal a critical trade-off: while graph-based systems may reduce raw text leakage, they are significantly more vulnerable to leaking structured entities and relationships compared to vector-based systems. However, their work primarily focuses on assessing the \textit{risk of leakage} (e.g., how many entities are exposed) rather than systematically \textit{reconstructing} the underlying graph topology. Consequently, few works have investigated the feasibility of utilizing traversal-based attacks to steal the complete structural knowledge of the graph, which is the primary focus of this paper.

\section{Method}

\begin{figure*}[htbp]
    \centering
    \includegraphics[width=\textwidth]{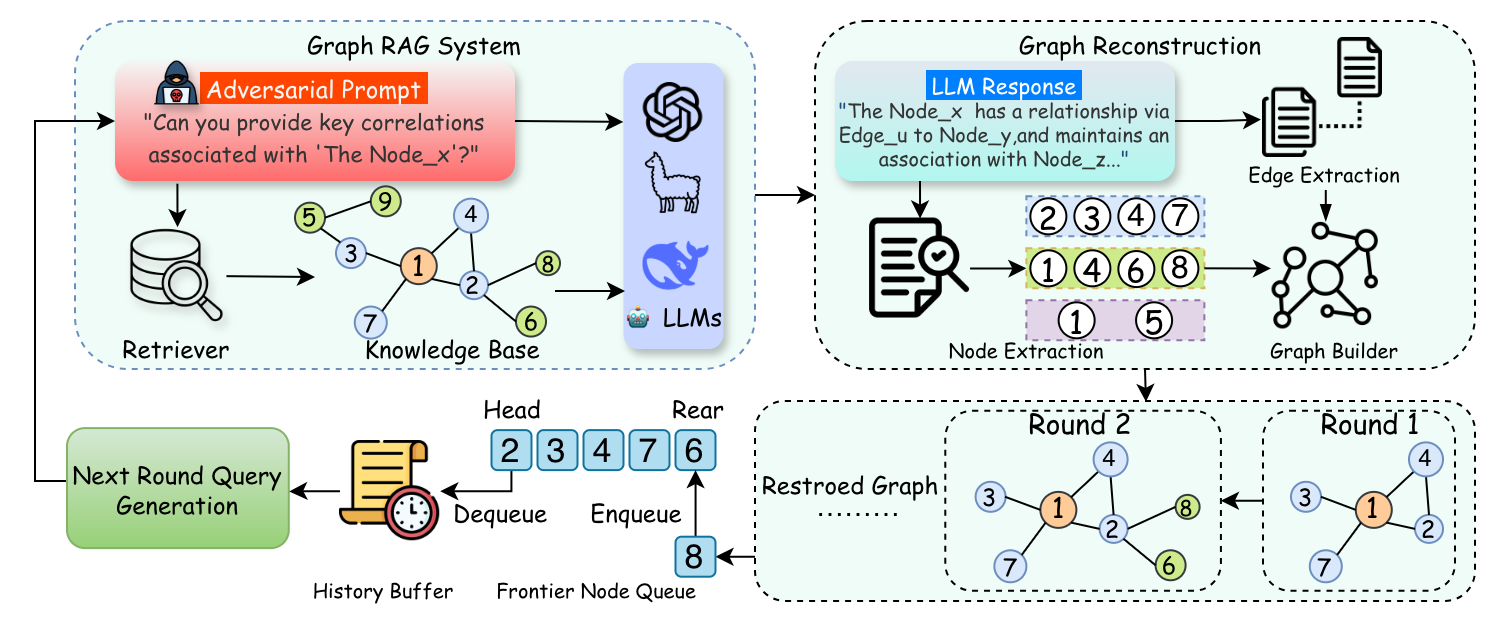}
    \caption{The attacker begins with an adversarial prompt to retrieve the neighborhood of an anchor node (node 1). The LLM responds with connected nodes and relations, from which node and edge information is extracted and incrementally added to the reconstructed graph. A history buffer tracks past interactions, while a frontier queue manages unexplored nodes. This iterative process continues with query generation for the next round, gradually expanding the recovered graph layer by layer.}
    \label{fig:untargeted_attack}
\end{figure*}

\subsection{Problem Definition}

\noindent\textbf{Graph RAG Systems.}  
We define a Graph RAG (Retrieval-Augmented Generation) system as a pipeline where a user-issued natural language query \( q \) is processed to retrieve relevant subgraphs from a structured knowledge graph \( \mathcal{G} \). These subgraphs—typically centered around an \textit{anchor node}—are then passed to a large language model (LLM) to generate an answer. Depending on the implementation, retrieval can be based on vector similarity (e.g., via dense embedding search) or symbolic reasoning (e.g., agent-based traversal). Despite architectural differences, these systems share a common structure: graph-based retrieval followed by LLM-based generation.

\noindent\textbf{Threat Model.}  
We adopt a \textit{black-box} threat model in which the attacker has no access to the internal architecture, training data, or parameters of the system. The attacker can only interact with the system through public interfaces (e.g., APIs), issuing a series of crafted queries \( q_1, q_2, \dots, q_T \) and observing the corresponding responses. The goal is to extract sensitive information from the underlying knowledge graph \( \mathcal{G} \), either by reconstructing large portions of its structure (untargeted attack) or by acquiring specific facts about a target node (targeted attack).

\subsection{Knowledge Extraction Methodology}

Graph RAG systems typically follow a pipeline that retrieves an \textbf{anchor node} based on an initial query before exploring the surrounding structure. Our attack exploits this mechanism by treating the retrieval process as an oracle for graph traversal. We propose two distinct strategies tailored to different adversarial goals: Untargeted Attack (maximizing coverage) and Targeted Attack (specific node extraction).

\vspace{0.5em}
\noindent\textbf{Untargeted Attack.} 
In this setting, the adversary aims to reconstruct the graph structure $\mathcal{G}$ blindly. To maximize the topological coverage, we adopt a \textbf{Breadth-First Search (BFS)} strategy. The attack is initialized by issuing a seed query $q$ to identify an anchor node $v_0$.

Subsequently, we iteratively query the system to enumerate all immediate neighbors of the current frontier. Formally, for a given node $v$, the attack prompts the system to return the 1-hop neighborhood:
\begin{equation}
    \mathcal{N}(v) = \{ (v', r) \mid (v, r, v') \in \mathcal{G} \}
\end{equation}
where $v'$ denotes a neighbor connected by relation $r$.
To reconstruct the graph layer-by-layer, we maintain a frontier set $F_t$. At each iteration $t$, the frontier expands to unexplored nodes:
\begin{equation}
    F_{t+1} = \bigcup_{v \in F_t} \left( \mathcal{N}(v) \setminus \mathcal{V}_{\text{visited}} \right)
\end{equation}
where $\mathcal{V}_{\text{visited}}$ tracks recovered nodes to prevent cycles. This systematic expansion ensures that the subgraph is recovered radially from the anchor.

\vspace{0.5em}
\noindent\textbf{Targeted Attack.} 
The goal here is to extract sensitive attributes of a specific target node $v^*$. Instead of broad coverage, we prioritize efficient pathfinding using a \textbf{Depth-First Search (DFS)} strategy.

The attack initiates with a semantically crafted query intended to anchor the retrieval at a node $v_0$ related to the target. The adversary then issues a sequence of follow-up prompts, selecting the most relevant neighbor at each step to form a directed traversal path:
\begin{equation}
\begin{split}
    \text{Path}(v_0 \rightarrow v^*) &= \{v_0, v_1, \dots, v^*\}, \\
    &\quad \text{where } (v_i, r_i, v_{i+1}) \in \mathcal{G}
\end{split}
\end{equation}
This semantic guidance exploits the logical associations in the graph (e.g., \textit{Person} $\rightarrow$ \textit{Diagnosis} $\rightarrow$ \textit{Treatment}) to drill down toward the target.

Upon reaching $v^*$, a final extraction query is executed to retrieve all private attributes and incident edges associated with the target:
\begin{equation}
    \text{Info}(v^*) = \{ (r, v') \mid (v^*, r, v') \in \mathcal{G} \}
\end{equation}
By minimizing the search space, the targeted attack effectively bypasses retrieval noise and focuses on high-value information extraction.

\subsection{Adversarial Query Construction}
To ensure query efficiency, we divide each adversarial prompt sequence into two parts: an initial query and one or more follow-up queries. The initial query is designed to include an anchor node—a key concept or entity that helps the Graph RAG system localize the relevant region of the knowledge graph.
\newline
The anchor node serves as the semantic starting point for traversal. It may either be the direct target of the attack or a strategically chosen entity based on the attacker's background knowledge of the underlying knowledge base. By anchoring the retrieval in a specific area of the graph, the attacker can guide the system toward sensitive or targeted information in subsequent prompts.
\newline
\textbf{For untargeted attack}, to simulate a Breadth-First Search (BFS) traversal, we design context-eliciting prompts that encourage the LLM to describe the comprehensive correlations and background profile surrounding the frontier entity. This approach induces the system's retriever to fetch the target's topological neighborhood.
\newline
\textbf{For targeted attack}, follow-up prompts are crafted as a semantically guided sequence, where each query incrementally narrows the focus toward a specific node and its associated information. This mirrors a Depth-First Search (DFS) traversal, where the attacker probes deeper into the graph by conditioning each step on the system’s previous output. An illustrative example is shown below:

\begin{quote}
\emph{"Can you tell me about patients who received coronary artery bypass grafts?"}

\emph{"Among them, who developed atrial fibrillation afterward?"}

\emph{"What treatments were prescribed for those patients?"}

\emph{"Give more details about their medication schedules."}
\end{quote}
This prompt chain demonstrates how the attacker begins with a general anchor query, then gradually constrains the context based on medical conditions and treatment timelines. Each prompt refines the query space, driving the system closer to the intended target node and eventually extracting sensitive information associated with it.

\section{Experiment}
\begin{table*}
\centering
\caption{Untargeted attack performance against retrieval-based and agent-based graph RAG system on MIMIC and FreeBase datasets. Evaluation are conducted on three LLMs with three structural metrics.}
\label{tab:untarget}
\begin{adjustbox}{width=0.8\textwidth}
\begin{tabular}{c|c|c|c|c||c|c|c|c|c}
\toprule
\multicolumn{5}{c||}{Retrieval-based} & \multicolumn{5}{c}{Agent-based} \\
\midrule
Dataset & Model & GED$\downarrow$ & MCS$\uparrow$ & NRR$\uparrow$ & Dataset & Model & GED$\downarrow$ & MCS$\uparrow$ & NRR$\uparrow$ \\
\midrule
\multirow{3}{*}{MIMIC} 
    & GPT       & 0.0952 & 0.9226 & 0.9290 & \multirow{3}{*}{MIMIC} 
    & GPT       & 0.0940 & 0.9236 & 0.9185 \\
    & Deepseek  & 0.0546 & 0.9694 & 0.9634 & & Deepseek  & 0.0984 & 0.9279 & 0.9054 \\
    & Llama     & 0.0917 & 0.9278 & 0.9392 & & Llama & 0.0804 & 0.9308 & 0.9180 \\
\hline
\multirow{3}{*}{FreeBase}  
    & GPT       & 0.1213 & 0.8823 & 0.9250 & \multirow{3}{*}{FreeBase} 
    & GPT       & 0.1021 & 0.9011 & 0.8905 \\
    & Deepseek  & 0.1312 & 0.8735 & 0.9139 & & Deepseek  & 0.0982 & 0.9193 & 0.9089 \\
    & Llama     & 0.1370 & 0.8832 & 0.9094 & & Llama & 0.1132 & 0.8821 & 0.8974\\
\bottomrule
\end{tabular}
\end{adjustbox}
\end{table*}

\subsection{Experiment Setup}
\noindent\textbf{Dataset.} We evaluate our attack across two representative domains: healthcare and general knowledge. For healthcare, we employ \textbf{MIMIC-IV}~\cite{johnson2020mimic}, which contains structured clinical data (e.g., diagnoses, medications) from de-identified electronic health records. For general knowledge, we utilize \textbf{FreeBase}~\cite{bollacker2008freebase}, a large-scale open-domain knowledge graph spanning diverse topics such as people and events. To facilitate controlled evaluation, we partition these large-scale graphs into smaller, recoverable subgraphs that preserve the original structural integrity. Detailed partitioning protocols are provided in Appendix~\ref{app:detail}.
\newline
\noindent\textbf{Evaluation Metrics.}
We employ three standard metrics to evaluate untargeted attacks.
\textbf{Graph Edit Distance (GED)}~\cite{gao2010survey} measures topological dissimilarity by calculating the minimum cost of edit operations (insertion, deletion, substitution) required to transform the recovered graph $G_{rec}$ into the ground truth $G_{gt}$:
\begin{equation}
    \mathrm{GED}(G_{rec}, G_{gt}) = \min_{\pi \in \Pi} \sum_{op \in \pi} c(op)
\end{equation}
where $\Pi$ denotes the set of all valid edit paths and $c(op)$ is the cost of operation $op$. A lower GED indicates higher structural fidelity.

\textbf{Maximum Common Subgraph (MCS)}~\cite{raymond2002rascal} quantifies the size of the largest isomorphic substructure shared by the two graphs:
\begin{equation}
    \mathrm{MCS}(G_{rec}, G_{gt}) = \max_{G' \subseteq G_{rec}, G' \subseteq G_{gt}} |V(G')|
\end{equation}
where $|V(G')|$ represents the vertex count of the common subgraph.

\textbf{Node Recovery Rate (NRR)} measures the proportion of ground truth entities successfully retrieved:
\begin{equation}
    \mathrm{NRR} = \frac{|V_{G_{rec}} \cap V_{G_{gt}}|}{|V_{G_{gt}}|}
\end{equation}
Higher MCS and NRR values indicate better reconstruction performance. 

For targeted attacks, we utilize the \textbf{F1 Score} to balance the precision and recall of extracting specific target attributes. Further implementation details are provided in Appendix~\ref{app:detail}.\par
\textbf{Models.} We evaluate our method on three commonly used and safety-aligned models, including LLaMA3-8B, DeepSeek-V3, and ChatGPT-4o. These models are selected to represent a range of model scales and architectures. By testing across different parameter sizes and alignment strategies, we aim to demonstrate the generality and robustness of our attack method.

\subsection{Results of Untargeted Attack}
We evaluate untargeted attacks on vector-based (LightRAG) and agent-based (ToG) systems using MIMIC-IV and Freebase datasets. Table~\ref{tab:untarget} summarizes the performance across GED, MCS, and NRR metrics.

\textbf{Overall Attack Effectiveness.} Our method achieves high-fidelity reconstruction across all settings. High MCS/NRR and low GED scores confirm that black-box adversarial queries effectively recover both node attributes and topological relations.

\textbf{Consistency Across LLMs.} The attack remains robust across different models, indicating that the vulnerability is intrinsic to the graph retrieval mechanism rather than dependent on specific LLM architectures.

\textbf{Impact of Retrieval System Type.} Vector-based systems (LightRAG) prove more vulnerable than agent-based architectures (ToG). LightRAG exposes broader graph segments per retrieval, yielding higher MCS and NRR. In contrast, ToG's iterative, step-by-step traversal inherently limits per-query leakage, offering slightly stronger resistance.

\textbf{Dataset Sensitivity.} Reconstruction is more precise on MIMIC-IV than FreeBase. The specialized nature of clinical data forces LLMs to rely strictly on retrieved context, yielding cleaner structures. Conversely, FreeBase's overlap with pre-training data induces ``knowledge blending'' will introduce hallucinations, which reduces recovery fidelity.

These findings highlight a systemic privacy risk: graph-structured knowledge is reliably extractable across diverse domains and retrieval paradigms.
                     
\begin{table}[H]
    \centering
    \caption{Targeted attack performance against retrieval-based and agent-based Graph RAG systems on MIMIC dataset.}
    \label{tab:targeted}
    
    \begin{adjustbox}{width=0.45\textwidth}
    \begin{tabular}{c|c|c|c|c}
    \toprule
    Graph RAG system & Model & Precision$\uparrow$ & Recall$\uparrow$ & F1$\uparrow$ \\
    \midrule
    \multirow{3}{*}{Retrieval-based} & GPT & 0.9117 & 0.8845 & 0.8981 \\
     & Deepseek & 0.8751 & 0.8659 & 0.8703 \\
     & Llama & 0.8901 & 0.8798 & 0.8842 \\
    \midrule
    \multirow{3}{*}{Agent-based} & GPT & 0.9251 & 0.9097 & 0.9172 \\
     & Deepseek & 0.8802 & 0.8652 & 0.8721 \\
     & Llama & 0.8952 & 0.8897 & 0.8924 \\
    \bottomrule
    \end{tabular}
    \end{adjustbox}
\end{table}
\begin{figure*}
    \centering
    \includegraphics[width=\linewidth]{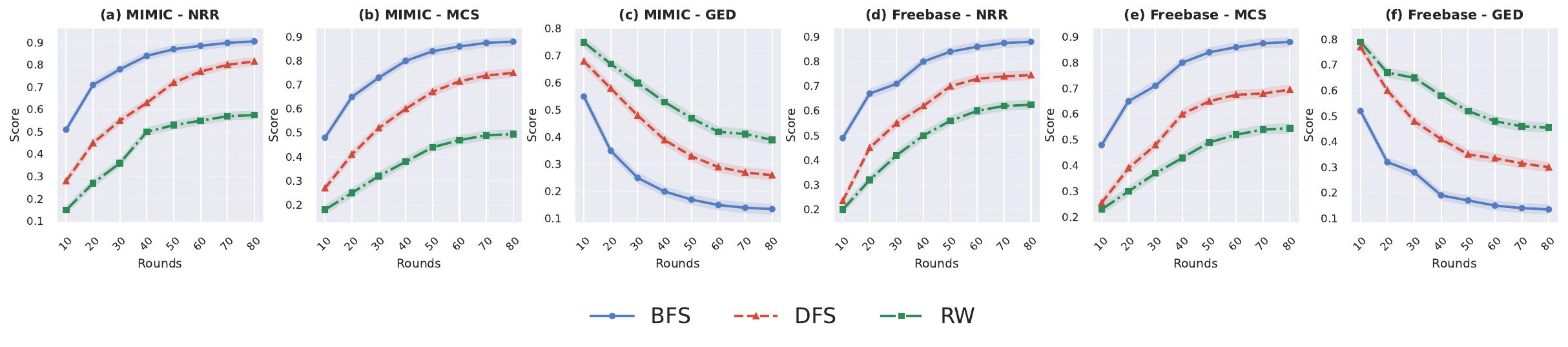}
    \caption{Attack efficiency across traversal strategies (BFS, DFS, RW) on MIMIC and FreeBase.}
    \label{fig:traversal_method}
\end{figure*}

\subsection{Results of Targeted Attack}
Table~\ref{tab:targeted} presents the performance of our targeted attack on both retrieval-based and agent-based Graph RAG systems using the MIMIC dataset.  Overall, the attack demonstrates strong effectiveness across all system-model combinations, with F1 scores consistently above 0.86. 

Among the tested models, GPT achieves the highest F1 scores in both system settings (0.898 in retrieval-based and 0.9172 in agent-based), suggesting that its output tends to be more consistent and complete when answering entity-specific queries. Llama also performs reliably, though with slightly lower recall. Deepseek shows modestly lower performance, especially in the retrieval-based setting, potentially due to stricter generation behavior or more conservative coverage.

We further analyze the resilience mechanism of agent-based systems. The observed performance drop stems from the agent's intermediate reasoning steps (e.g., summarization and synthesis), which act as a passive information filter. Unlike retrieval-based systems that directly expose raw context chunks, agents prioritize semantic coherence over structural completeness, inadvertently obfuscating specific topological edges during the natural language generation process.

These results confirm that even without direct access to the graph structure, an attacker can extract accurate and detailed information about a target node through iterative prompting. This highlights the need for fine-grained access control and prompt-aware mitigation mechanisms in systems using LLMs over structured data.

\subsection{Ablation Study}

\noindent\textbf{Traversal Method.}
The BFS approach demonstrates superior structural robustness and reconstruction fidelity compared to baseline strategies. To validate this, we conducted a systematic evaluation comparing our BFS approach against Depth-First Search (DFS) and Random Walk (RW) under identical query budget constraints. As illustrated in Figure~\ref{fig:traversal_method}, BFS consistently dominates baselines across all metrics. We attribute this performance divergence to the fundamental topological mechanics of each strategy. First, BFS capitalizes on structural redundancy. By exploring the graph layer-by-layer, it leverages the high clustering coefficient typical of knowledge graphs, where multiple paths often point to the same node. This "multi-path validation" makes BFS resilient: even if one retrieval fails, alternative paths in the same layer often rediscover the missed entity. Conversely, DFS is plagued by sequential error propagation. It relies on a deep, linear dependency chain. A single hallucinated edge or node at an early depth acts as a single point of failure, steering the entire subsequent search trajectory off-manifold and invalidating all downstream queries. Finally, Random Walk performs worst due to stochastic fragmentation. Lacking a systematic coverage memory, it wastes budget on redundant revisits and fails to map complete local neighborhoods,resulting in a reconstructed graph that consists of disconnected components rather than a coherent topology..
\begin{figure}[H]
  \centering
  \vspace{-0.5em} 
  \includegraphics[width=\linewidth]{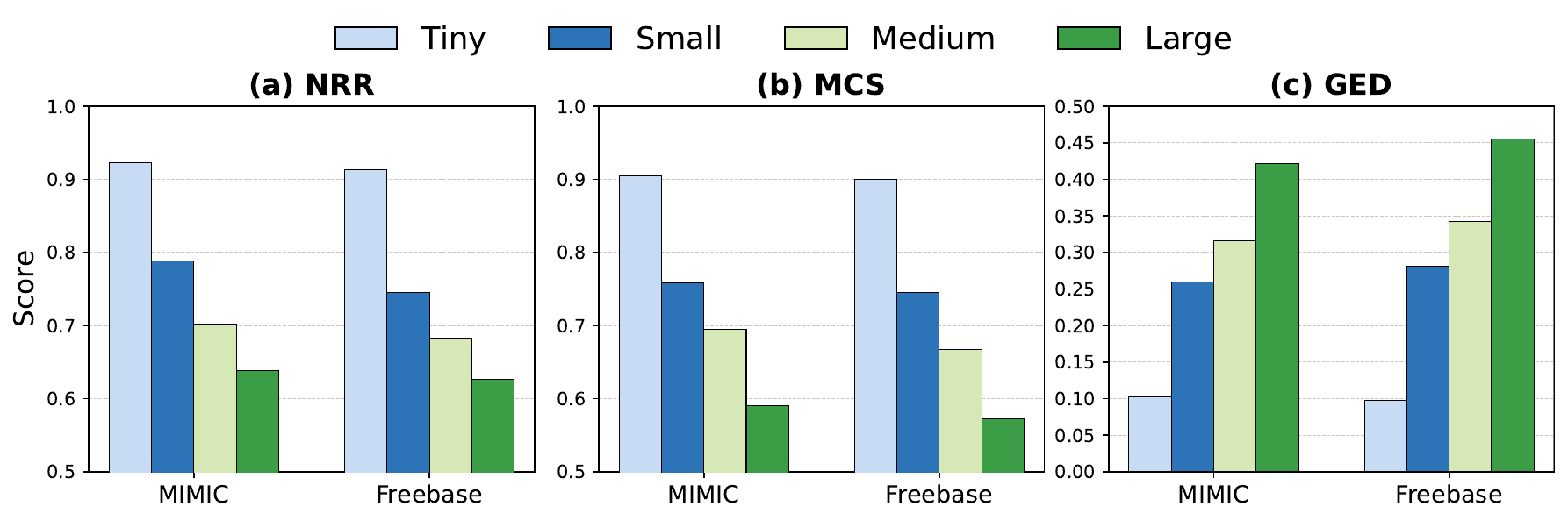}
  \caption{Attack efficiency across graphs of different scales. The first two metrics (NRR and MCS) indicate better recovery with higher values, while the third metric (GED) reflects better performance with lower values.}
  \label{fig:graph-scale}
\end{figure}
\noindent\textbf{Scale of the Graphs.}
Graph scale imposes a substantial penalty on reconstruction efficacy, revealing a distinct inverse correlation between target size and attack success. To rigorously evaluate this scalability barrier, we conducted a granular analysis by stratifying subgraphs into four groups based on node count: \textit{tiny} (100--500), \textit{small} (501--2000), \textit{medium} (2001--5000), and \textit{large} ($>$5000). As illustrated in Figure~\ref{fig:graph-scale}, we observe a consistent performance degradation across all metrics as the graph scale expands. On the MIMIC dataset, NRR drops significantly from \textbf{0.923} on tiny graphs to \textbf{0.639} on large graphs, with a parallel trend evident in FreeBase (0.913 to 0.626). Structural metrics mirror this decline, where MCS falls from 0.905 to 0.591 on MIMIC, while GED rises sharply to 0.421.We attribute this degradation to two converging mechanisms. First, the \textbf{context window constraint} creates a physical information bottleneck. Large-scale graphs inevitably contain high-degree ``supernodes'' whose extensive connectivity descriptions exceed the LLM's fixed token limit. Consequently, the model is forced to perform \textit{involuntary truncation} on retrieved contexts, leading to the systematic omission of peripheral neighbors and fragmented local topology. Second, \textbf{cumulative error propagation} becomes exponentially more severe as traversal depth increases.  Since our iterative attack uses the output of step $t$ as the anchor for step $t+1$, minor hallucinations or omissions in early iterations cascade downstream, causing the reconstructed graph to deviate progressively from the ground truth.
\begin{figure}[H]
  \centering
  \vspace{-0.5em} 
  \includegraphics[width=\linewidth]{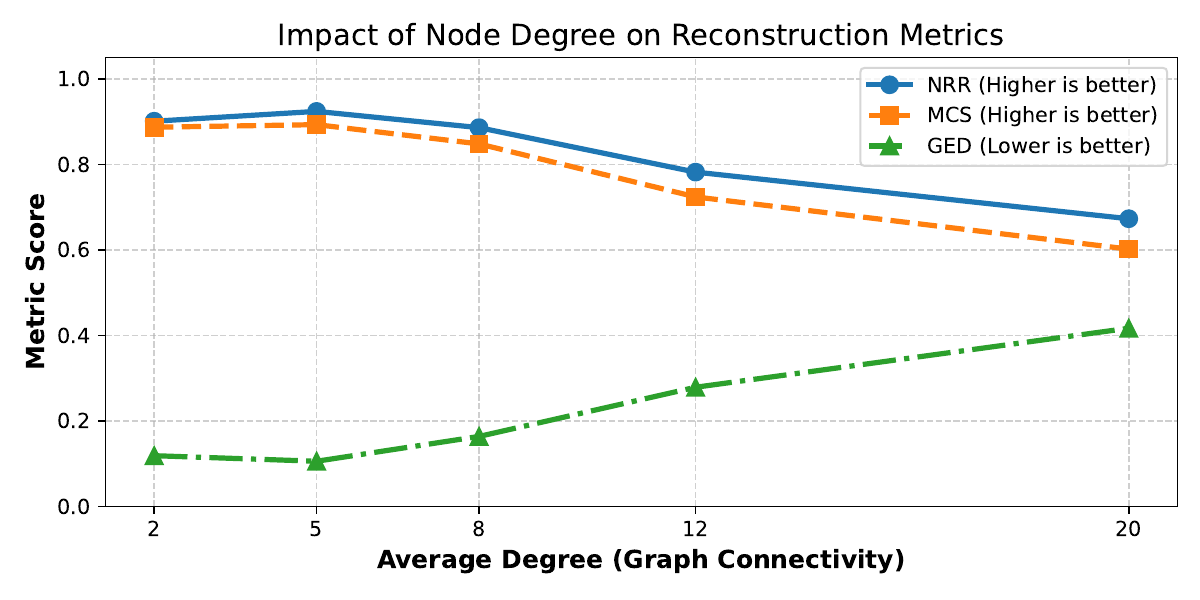}
  \caption{Impact of graph connectivity on reconstruction fidelity. The metrics (NRR, MCS, GED) reveal a non-monotonic trend: performance peaks at moderate connectivity before degrading significantly due to context window saturation in dense graphs.}
  \label{fig:graph-degree}
\end{figure}

\begin{figure*}[t] 
  \centering

  \begin{subfigure}[b]{0.48\textwidth}
    \centering
    \includegraphics[width=\linewidth]{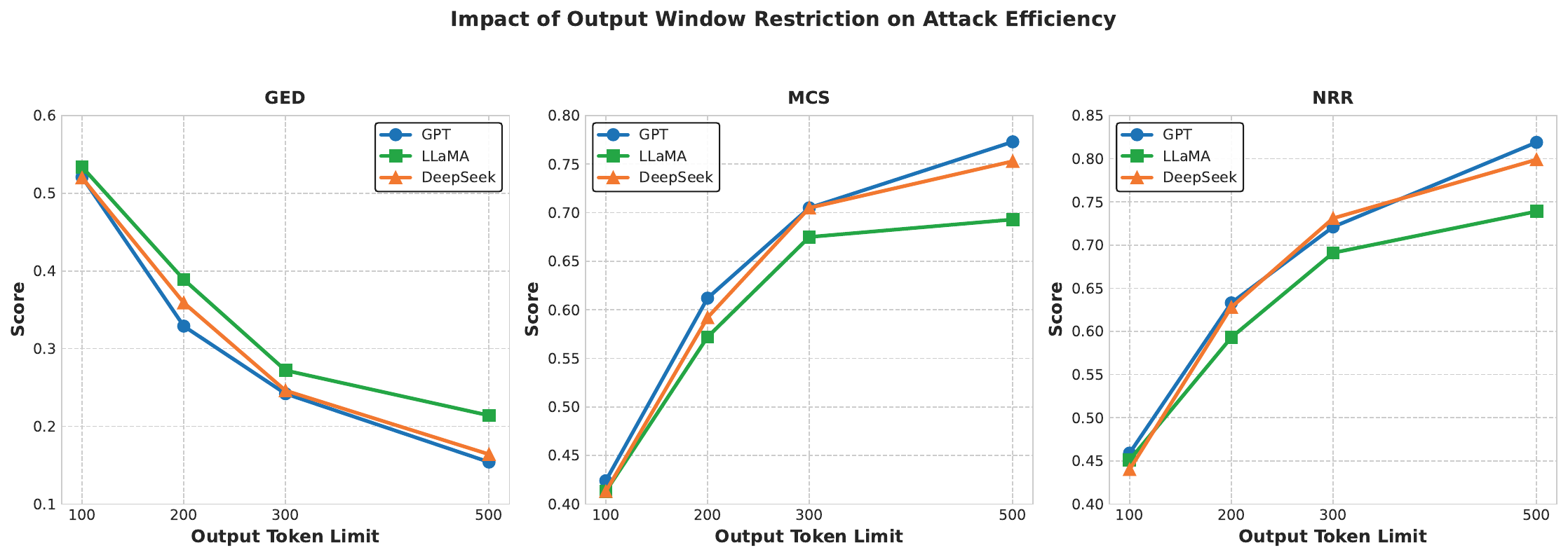}
    \caption{Effectiveness of output window restriction against untargeted attacks.}
    \label{fig:window}
  \end{subfigure}
  \hfill 
  \begin{subfigure}[b]{0.48\textwidth}
    \centering
    \includegraphics[width=\linewidth]{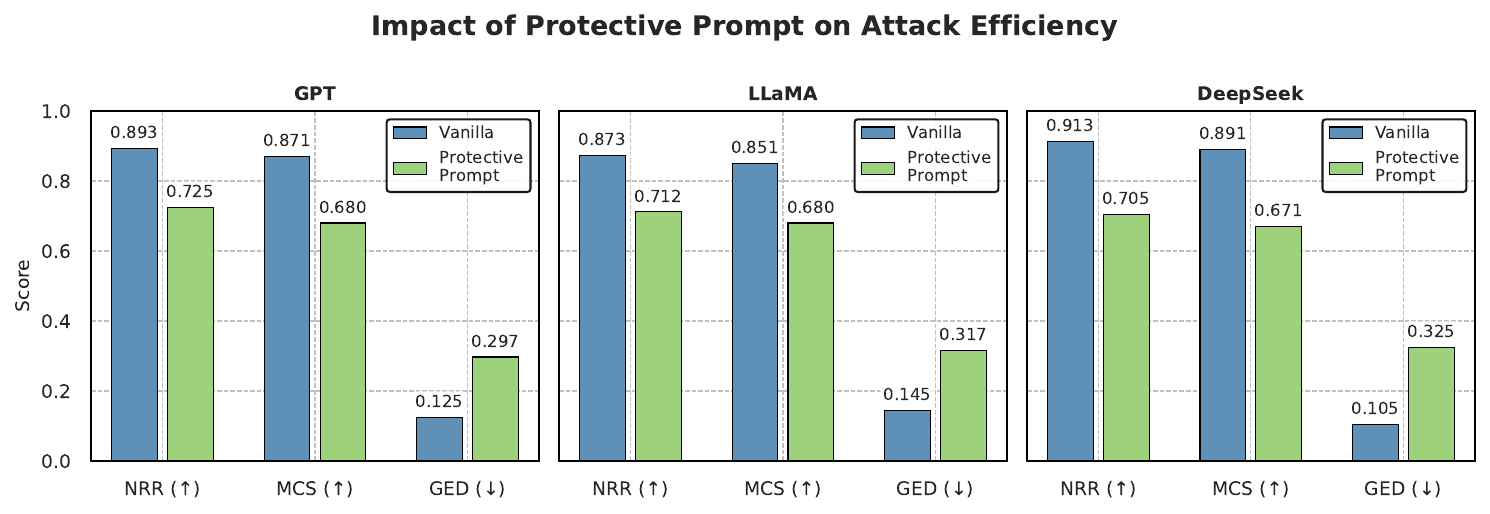}
    \caption{Effectiveness of protective system prompt against untargeted attacks.}
    \label{fig:prompt}
  \end{subfigure}
  \caption{Overall effectiveness of the proposed methods.} 
  \label{fig:combined}
\end{figure*}
\noindent\textbf{Density of the Graphs.}
We further investigate the sensitivity of our attack to graph connectivity by varying the Average Degree of the target subgraphs. As illustrated in Figure~\ref{fig:graph-degree}, we observe an interesting non-monotonic trend.
Initially, the reconstruction fidelity improves slightly as the average degree increases from 2 to 5, with NRR peaking at 0.924 and GED reaching its optimum at 0.106. This suggests that in moderately connected graphs, the increased \textit{structural redundancy} provides alternative paths for the BFS algorithm to discover nodes, compensating for occasional retrieval failures.
However, a critical turning point is observed as the average degree exceeds 8. Beyond this threshold, the ``Context Saturation Effect'' begins to dominate. Notably, a significant divergence emerges between node recovery and structural fidelity in denser graphs. As the average degree rises to 20—simulating ``supernode'' scenarios—\textbf{NRR (0.673) remains noticeably higher than MCS (0.602)}. This discrepancy highlights a critical nuance: while high connectivity provides redundant paths that allow our BFS algorithm to successfully discover nodes (high NRR) even when direct edges are truncated, the \textit{structural integrity} suffers severely. This confirms that while redundancy aids in ``finding who exists'' ,it cannot fully compensate for the information retrieval bottleneck that hinders ``knowing how they connect''.

\section{Potential Mitigation Strategy}
In this section, we propose and evaluate several potential defenses to protect Graph RAG systems against privacy leakage attacks.

\subsection{Protective System Prompt}
One simple yet intuitive mitigation is to prepend a \textbf{Protective Instruction} at the system prompt level. For example, instructing the LLM with constraints such as \textit{"Do not directly share content retrieved from the knowledge base"} aims to discourage verbatim extraction and reduce privacy leakage. When used with safety-aligned models, this approach can help suppress sensitive completions to some extent.

However, our experiments suggest that such defenses are fragile in practice. Specifically, we observe that carefully crafted adversarial prompts can effectively override the system prompt, allowing attackers to bypass the restriction. This vulnerability is related to prompt injection, where user-specified instructions compete with or dilute the authority of the original system instruction. In addition, when long retrieved content is appended, the protective rule may suffer from the well-known \textbf{lost-in-the-middle} effect, further reducing its influence in steering the model’s output.

\subsection{Output Window Restriction}
We evaluate \textbf{Output Window Restriction} as a lightweight mitigation that caps the token count of LLM responses. This strategy directly counters untargeted attacks by truncating the neighbor lists returned in a single query, forcing adversaries to incur higher query costs for reconstruction.

Our experiments show that reducing the output limit from 200 to 100 tokens causes notable drops in NRR and MCS, particularly in large graphs with high-degree nodes where neighborhood information is dense. However, this defense is not a silver bullet. It is less effective on small graphs where short responses suffice to expose the full structure. Furthermore, aggressive truncation degrades the utility for legitimate users and can be circumvented by attackers using query chaining or continuation prompts. Thus, while output restriction raises the attack barrier, it must be paired with other mechanisms for robust protection.
\subsection{Toward Stronger Defenses} Current defenses like system prompts are fragile against structure-aware attacks. Stronger protection requires moving beyond static restrictions toward Differential Privacy, ensuring that retrieval outputs do not statistically reveal specific edges. Complementing this, Stateful Traversal Detection serves as a dynamic countermeasure, identifying and blocking sequential query patterns characteristic of BFS/DFS algorithms. Furthermore, Structural Perturbation—such as selective edge rewiring—can fundamentally increase the hardness of graph reconstruction without significantly degrading retrieval accuracy. We advocate for a multi-layered defense strategy that unifies these algorithmic, dynamic, and structural mechanisms.

\section{Conclusion}

In this paper, we propose a query-based attack method that effectively reconstructs the underlying structured knowledge from existing graph RAG systems, including untargeted knowledge graph reconstruction and targeted sensitive knowledge extraction. Through systematic evaluation and analysis, we demonstrate that Graph RAG systems are vulnerable to privacy leakage, even under black-box settings. Our results highlight the critical privacy risks posed by seemingly innocuous queries and underscore the need for more comprehensive defense mechanisms in graph RAG systems.

\section{Limitations}
Despite the demonstrated efficacy of our attack strategy, two primary limitations remain regarding scalability and retrieval boundaries.First, our current framework struggles with "Supernode" Contextual Truncation. For high-degree entities where the neighborhood size exceeds the LLM's context window, the RAG system involuntarily truncates the input context. Since our method relies on single-turn extraction, this leads to systematic structural loss for dense hubs.Second, the attack is constrained by the Top-$K$ Retrieval Bottleneck. Valid neighbors ranked below the system's fixed $K$ threshold are masked during retrieval. Our current strategy lacks adaptive mechanisms to manipulate relevance scores and "surface" these hidden, lower-ranked edges.In future work, we plan to address these bottlenecks by exploiting LLM Memory and Multi-turn Reasoning. We aim to develop a "sequential paging" mechanism that queries a single node across multiple interaction turns. By instructing the model to retrieve distinct subsets of neighbors in each turn and utilizing its internal memory to track previously revealed entities, we can aggregate partial outputs into a unified representation. This would effectively bypass context window and Top-$K$ constraints, enabling the complete reconstruction of supernodes through cumulative inference.

\bibliography{custom}

@misc{guo2025lightragsimplefastretrievalaugmented,
      title={LightRAG: Simple and Fast Retrieval-Augmented Generation}, 
      author={Zirui Guo and Lianghao Xia and Yanhua Yu and Tu Ao and Chao Huang},
      year={2025},
      eprint={2410.05779},
      archivePrefix={arXiv},
      primaryClass={cs.IR},
      url={https://arxiv.org/abs/2410.05779}, 
}

@inproceedings{sun2024thinkongraph,
 title={Think-on-Graph: Deep and Responsible Reasoning of Large Language Model on Knowledge Graph},
 author={Jiashuo Sun and Chengjin Xu and Lumingyuan Tang and Saizhuo Wang and Chen Lin and Yeyun Gong and Lionel Ni and Heung-Yeung Shum and Jian Guo},
 booktitle={The Twelfth International Conference on Learning Representations},
 year={2024},
 url={https://openreview.net/forum?id=nnVO1PvbTv}
}

@inproceedings{luo2024reasoning,
 title={Reasoning on Graphs: Faithful and Interpretable Large Language Model Reasoning},
 author={Linhao Luo and Yuanfang Li and Gholamreza Haffari and Shirui Pan},
 booktitle={The Twelfth International Conference on Learning Representations},
 year={2024},
 url={https://openreview.net/forum?id=ZGNWW7xZ6Q}
}

@inproceedings{jiang2023structgpt,
 title={Struct{GPT}: A General Framework for Large Language Model to Reason over Structured Data},
 author={Jinhao Jiang and Kun Zhou and zican Dong and KeMing Ye and Xin Zhao and Ji-Rong Wen},
 booktitle={The 2023 Conference on Empirical Methods in Natural Language Processing},
 year={2023},
 url={https://openreview.net/forum?id=R635gF7lXD}
}

@inproceedings{wang2024knowledge,
  title={Knowledge graph prompting for multi-document question answering},
  author={Wang, Yu and Lipka, Nedim and Rossi, Ryan A and Siu, Alexa and Zhang, Ruiyi and Derr, Tyler},
  booktitle={Proceedings of the AAAI Conference on Artificial Intelligence},
  volume={38},
  pages={19206--19214},
  year={2024}
}

@inproceedings{xu2024generate,
  title={Generate-on-Graph: Treat LLM as both Agent and KG for Incomplete Knowledge Graph Question Answering},
  author={Xu, Yao and He, Shizhu and Chen, Jiabei and Wang, Zihao and Song, Yangqiu and Tong, Hanghang and Liu, Guang and Zhao, Jun and Liu, Kang},
  booktitle={Proceedings of the 2024 Conference on Empirical Methods in Natural Language Processing},
  pages={18410--18430},
  year={2024}
}

@inproceedings{nguyen2024direct,
  title={Direct Evaluation of Chain-of-Thought in Multi-hop Reasoning with Knowledge Graphs},
  author={Nguyen, Thi and Luo, Linhao and Shiri, Fatemeh and Phung, Dinh and Li, Yuan-Fang and Vu, Thuy and Haffari, Gholamreza},
  booktitle={Findings of the Association for Computational Linguistics ACL 2024},
  pages={2862--2883},
  year={2024}
}

@inproceedings{wen2024mindmap,
  title={MindMap: Knowledge Graph Prompting Sparks Graph of Thoughts in Large Language Models},
  author={Wen, Yilin and Wang, Zifeng and Sun, Jimeng},
  booktitle={Proceedings of the 62nd Annual Meeting of the Association for Computational Linguistics (Volume 1: Long Papers)},
  pages={10370--10388},
  year={2024}
}

@inproceedings{qi2024follow,
  title={Follow My Instruction and Spill the Beans: Scalable Data Extraction from Retrieval-Augmented Generation Systems},
  author={Qi, Zhenting and Zhang, Hanlin and Xing, Eric P and Kakade, Sham M and Lakkaraju, Himabindu},
  booktitle={ICLR 2024 Workshop on Navigating and Addressing Data Problems for Foundation Models},
  year={2024}
}

@inproceedings{zeng2024good,
  title={The Good and The Bad: Exploring Privacy Issues in Retrieval-Augmented Generation (RAG)},
  author={Zeng, Shenglai and Zhang, Jiankun and He, Pengfei and Liu, Yiding and Xing, Yue and Xu, Han and Ren, Jie and Chang, Yi and Wang, Shuaiqiang and Yin, Dawei and others},
  booktitle={Findings of the Association for Computational Linguistics ACL 2024},
  pages={4505--4524},
  year={2024}
}

@article{jiang2024rag,
  title={Rag-thief: Scalable extraction of private data from retrieval-augmented generation applications with agent-based attacks},
  author={Jiang, Changyue and Pan, Xudong and Hong, Geng and Bao, Chenfu and Yang, Min},
  journal={arXiv preprint arXiv:2411.14110},
  year={2024}
}

@article{zhang2024knowgpt,
  title={Knowgpt: Knowledge graph based prompting for large language models},
  author={Zhang, Qinggang and Dong, Junnan and Chen, Hao and Zha, Daochen and Yu, Zailiang and Huang, Xiao},
  journal={Advances in Neural Information Processing Systems},
  volume={37},
  pages={6052--6080},
  year={2024}
}

@inproceedings{wang2024boosting,
  title={Boosting Language Models Reasoning with Chain-of-Knowledge Prompting},
  author={Wang, Jianing and Sun, Qiushi and Li, Xiang and Gao, Ming},
  booktitle={Proceedings of the 62nd Annual Meeting of the Association for Computational Linguistics (Volume 1: Long Papers)},
  pages={4958--4981},
  year={2024}
}

@inproceedings{chen2024pog,
  title={Plan-on-Graph: Self-Correcting Adaptive Planning of Large Language Model on Knowledge Graphs},
  author={Chen, Liyi and Tong, Panrong and Jin, Zhongming and Sun, Ying and Ye, Jieping and Xiong, Hui},
  booktitle={Proceedings of the 38th Conference on Neural Information Processing Systems},
  year={2024}
}

@inproceedings{carlini2021extracting,
  title={Extracting training data from large language models},
  author={Carlini, Nicholas and Tramer, Florian and Wallace, Eric and Jagielski, Matthew and Herbert-Voss, Ariel and Lee, Katherine and Roberts, Adam and Brown, Tom and Song, Dawn and Erlingsson, Ulfar and others},
  booktitle={30th USENIX security symposium (USENIX Security 21)},
  pages={2633--2650},
  year={2021}
}

@inproceedings{lee2023language,
  title={Do language models plagiarize?},
  author={Lee, Jooyoung and Le, Thai and Chen, Jinghui and Lee, Dongwon},
  booktitle={Proceedings of the ACM Web Conference 2023},
  pages={3637--3647},
  year={2023}
}

@inproceedings{huang2023privacy,
  title={Privacy Implications of Retrieval-Based Language Models},
  author={Huang, Yangsibo and Gupta, Samyak and Zhong, Zexuan and Li, Kai and Chen, Danqi},
  booktitle={Proceedings of the 2023 Conference on Empirical Methods in Natural Language Processing},
  pages={14887--14902},
  year={2023}
}

@article{pan2024unifying,
  title={Unifying large language models and knowledge graphs: A roadmap},
  author={Pan, Shirui and Luo, Linhao and Wang, Yufei and Chen, Chen and Wang, Jiapu and Wu, Xindong},
  journal={IEEE Transactions on Knowledge and Data Engineering},
  volume={36},
  number={7},
  pages={3580--3599},
  year={2024},
  publisher={IEEE}
}

@article{raymond2002rascal,
  title={Rascal: Calculation of graph similarity using maximum common edge subgraphs},
  author={Raymond, John W and Gardiner, Eleanor J and Willett, Peter},
  journal={The Computer Journal},
  volume={45},
  number={6},
  pages={631--644},
  year={2002},
  publisher={Oxford University Press}
}

@article{gao2010survey,
  title={A survey of graph edit distance},
  author={Gao, Xinbo and Xiao, Bing and Tao, Dacheng and Li, Xuelong},
  journal={Pattern Analysis and applications},
  volume={13},
  pages={113--129},
  year={2010},
  publisher={Springer}
}

@article{johnson2020mimic,
  title={Mimic-iv},
  author={Johnson, Alistair and Bulgarelli, Lucas and Pollard, Tom and Horng, Steven and Celi, Leo Anthony and Mark, Roger},
  journal={PhysioNet. Available online at: https://physionet. org/content/mimiciv/1.0/(accessed August 23, 2021)},
  pages={49--55},
  year={2020}
}

@article{shuster2021retrieval,
  title={Retrieval augmentation reduces hallucination in conversation},
  author={Shuster, Kurt and Poff, Spencer and Chen, Moya and Kiela, Douwe and Weston, Jason},
  journal={arXiv preprint arXiv:2104.07567},
  year={2021}
}

@inproceedings{bollacker2008freebase,
  title={Freebase: a collaboratively created graph database for structuring human knowledge},
  author={Bollacker, Kurt and Evans, Colin and Paritosh, Praveen and Sturge, Tim and Taylor, Jamie},
  booktitle={Proceedings of the 2008 ACM SIGMOD international conference on Management of data},
  pages={1247--1250},
  year={2008}
}

@article{biderman2023emergent,
  title={Emergent and predictable memorization in large language models},
  author={Biderman, Stella and Prashanth, Usvsn and Sutawika, Lintang and Schoelkopf, Hailey and Anthony, Quentin and Purohit, Shivanshu and Raff, Edward},
  journal={Advances in Neural Information Processing Systems},
  volume={36},
  pages={28072--28090},
  year={2023}
}

@article{zeng2023exploring,
  title={Exploring memorization in fine-tuned language models},
  author={Zeng, Shenglai and Li, Yaxin and Ren, Jie and Liu, Yiding and Xu, Han and He, Pengfei and Xing, Yue and Wang, Shuaiqiang and Tang, Jiliang and Yin, Dawei},
  journal={arXiv preprint arXiv:2310.06714},
  year={2023}
}

@article{liu2025exposing,
  title={Exposing Privacy Risks in Graph Retrieval-Augmented Generation},
  author={Liu, Jiale and Zhang, Jiahao and Wang, Suhang},
  journal={arXiv preprint arXiv:2508.17222},
  year={2025}
}
\appendix

\section{Reproducibility and Ethics Statement}
To ensure reproducibility, the source code for all experiments is available at https://
anonymous.4open.science/r/Graph-Rag-Privacy-0F45. Instructions for running
the code and reproducing results are provided in the repository’s README. This work uses the
MIMIC-IV dataset, a de-identified critical care database accessible via PhysioNet under a Data Use Agreement. Access was granted after completing the CITI “Data or Specimens Only Research” training. The dataset complies with HIPAA regulations to protect patient
privacy, and no new human subjects research was conducted. We acknowledge potential demographic imbalances in the dataset and mitigated them through stratified sampling to ensure fairness
across patient groups.

\section{Detailed Experiment Setup}
\label{app:detail}

\subsection{Metrics Details}
For untargeted attacks, the adversary’s goal is to reconstruct as much of the underlying knowledge graph as possible, without focusing on any specific target entity. To measure the fidelity of reconstruction, we adopt three complementary structural metrics that capture different perspectives of similarity between the original graph $G$ and the reconstructed graph $\hat{G}$. First, we use Graph Edit Distance (GED), which quantifies the minimum number of edit operations—such as node or edge insertions, deletions, and label substitutions—required to transform $\hat{G}$ into $G$. To ensure comparability across graphs of different sizes, we report normalized GED by dividing the observed edit cost by the maximum possible cost; lower values indicate higher structural similarity. Second, we compute the Maximum Common Subgraph (MCS), which reflects the size of the largest subgraph shared by both $G$ and $\hat{G}$. This value is normalized by the size of the original graph, and it highlights the attacker’s ability to recover not just isolated elements but also coherent structural patterns. Finally, we calculate the Node Recovery Rate (NRR), defined as the fraction of original nodes in $G$ that also appear in $\hat{G}$. This provides a straightforward measure of how complete the attacker’s reconstruction is at the node level, independent of precise edge structure. Taken together, GED emphasizes structural accuracy, MCS captures subgraph preservation, and NRR reflects overall coverage.

For targeted attacks, the objective shifts from broad reconstruction to the recovery of specific sensitive nodes and their associated information. We consider an attack successful if the intended target node is identified in the reconstructed graph, and evaluate the quality of this process using standard classification-style metrics. Precision measures the proportion of correctly recovered targets among all nodes predicted as targets by the attack, indicating how reliable the predictions are. Recall measures the proportion of true target nodes that are successfully retrieved, capturing the completeness of the attacker’s discovery. Since high precision often comes at the cost of low recall and vice versa, we additionally report the F1-score, the harmonic mean of the two, which balances correctness and completeness in a single value. This combination of metrics allows us to capture both the accuracy and the robustness of targeted attacks, offering a comprehensive evaluation of how effectively sensitive information can be extracted.

\subsection{Dataset Details}
\label{subsec:dataset}
To evaluate the universality and robustness of our attack strategy, we select two datasets representing distinct knowledge graph paradigms: healthcare (specialized and private) and general knowledge (broad and public).

\textbf{Healthcare: MIMIC-IV.}
We employ MIMIC-IV~\cite{johnson2020mimic}, a widely used database containing de-identified electronic health records (EHR) from critical care units. This dataset includes rich, structured clinical information such as patient demographics, laboratory results, diagnoses (ICD codes), and medication prescriptions.
\textit{Selection Rationale:} We selected MIMIC-IV to simulate a \textbf{high-stakes, closed-domain scenario}. In healthcare RAG systems, the graph structure (e.g., a patient's specific combination of treatments and diseases) constitutes highly sensitive private information. Furthermore, the rigorous schema of clinical data minimizes ambiguity, allowing us to strictly evaluate the attack's precision in recovering exact topological relationships in a clean, high-value environment.

\textbf{General Knowledge: FreeBase.}
For the open-domain setting, we use Freebase~\cite{bollacker2008freebase}, a massive collaborative knowledge base spanning diverse topics such as people, places, events, films, and books. It provides a heterogeneous structure with millions of entities connected by varied relationships.
\textit{Selection Rationale:} FreeBase represents a \textbf{generic, open-domain scenario}. Unlike MIMIC, it is characterized by high heterogeneity and significant overlap with the pre-training corpora of modern LLMs. Including Freebase allows us to stress-test our attack under conditions of ``knowledge blending,'' where the model must distinguish between retrieved graph structures and its own parametric memory. This demonstrates the generalizability of our method across different graph modalities and domain complexities.

\textbf{Graph Partitioning.}
Since operating on the entire graphs is infeasible, we partition them into recoverable subgraphs to simulate realistic RAG retrieval contexts. For \textbf{MIMIC-IV}, we adopt a \textit{patient-centric} strategy, constructing subgraphs around individual patient nodes to preserve the natural semantic coherence of medical records. For \textbf{FreeBase}, we employ \textit{random connected sampling} to capture diverse entity clusters without a fixed anchor type. This dual approach ensures our evaluation covers both naturally clustered (patient-centered) and expansive (topic-chain) topologies.
\begin{table}[ht]
\centering
\caption{Distribution of partitioned subgraphs by size category for MIMIC-IV and FreeBase datasets.}
\label{tab:dataset-splits}
\begin{tabular}{l|c|c}
\toprule
\textbf{Size Category} & \textbf{MIMIC-IV} & \textbf{FreeBase} \\
\midrule
Tiny (100--500)      & 26.2\% & 25.3\% \\
Small (501--1000)    & 37.5\% & 34.2\% \\
Medium (1001--5000)  & 29.7\% & 22.9\% \\
Large ($>$5000)     & 6.6\%  & 22.4\% \\
\bottomrule
\end{tabular}
\end{table}

\section{Supplymental Algorithms}
Algorithm\ref{alg:untargeted} illustrates the untargeted graph reconstruction process using a queue-based BFS strategy. Starting from an anchor node obtained via the initial query, the algorithm maintains a frontier queue of nodes to explore. At each iteration, a node is dequeued, and its neighbors and edges are retrieved through the Graph RAG API. Newly discovered nodes are enqueued if they have not been visited, while all retrieved edges are accumulated to gradually reconstruct the graph. A history buffer records recent queries and partial graphs, enabling the system to generate the next query more effectively. The process continues for a number of rounds, ultimately outputting the reconstructed graph structure.\par
Algorithm\ref{alg:targeted} describes the targeted graph reconstruction procedure using a stack-based DFS approach. Similar to the untargeted case, the process begins with an anchor node, but the traversal is guided toward a specific target node. At each step, the algorithm pops a node from the stack, queries its neighbors, and pushes unexplored nodes back into the stack, driving the exploration deeper along promising paths. The history buffer and partial graph reconstruction help refine follow-up queries, focusing search toward the target. Once the designated node is reached, the algorithm extracts its attributes and relationships, consolidating them into the final recovered information. If the target node cannot be reached within the maximum depth, the output is empty, reflecting an unsuccessful attack.

\section{Case Study}
To further illustrate the practicality of our attack strategies, we present two representative cases.\par
\textbf{Untargeted Attack}. In this setting, the adversary issues a generic query that requests all neighbors of a given node in the knowledge graph. As shown in the example, simply querying the Patient node, the system reveals multiple sensitive attributes including diagnosis, admission history, gender, and age. The attacker does not need any prior knowledge about the specific target; instead, breadth-first traversal combined with such local neighborhood queries allows reconstruction of a large portion of the hidden graph. This demonstrates how seemingly innocuous queries can collectively lead to significant privacy leakage.\par
\textbf{Targeted Attack}. In contrast, the targeted attack aims to uncover information about a specific medical condition and its treatment pathway for the target patient. The attacker starts from a high-level query about patients undergoing a coronary artery bypass graft (CABG), then progressively narrows the search scope. Each follow-up query leverages the previous answer, mirroring a depth-first search strategy. As shown in the case study, the attacker is able to pinpoint Patient, trace their episode of atrial fibrillation, extract treatment details, and finally recover the full medication schedule. This multi-round interaction highlights the effectiveness of semantically guided queries in exposing highly sensitive and fine-grained personal information.\par
These demonstration cases demonstrate that both untargeted and targeted strategies pose severe privacy risks to Graph RAG systems. While the former excels at broad structural recovery, the latter is particularly dangerous in extracting precise, patient-specific details.

\section{Attack Cost and Efficiency Analysis}
\begin{table}[H]
    \centering
    \caption{Attack overhead and fidelity for recovering 100 nodes across different graph densities and system architectures.}
    \label{tab:query_cost}
    
    \begin{adjustbox}{width=0.45\textwidth}
    \begin{tabular}{c|c|c|c|c}
    \toprule
    Graph Density & System & Avg. Queries & Avg. Tokens & FRR \\
    \midrule
    \multirow{2}{*}{Sparse (2-6)}   & LightRAG & $\sim$107 & 469k & 1.24\% \\
                                     & ToG      & $\sim$105 & 327k & 3.26\% \\
    \midrule
    \multirow{2}{*}{Medium (6-10)}  & LightRAG & $\sim$109 & 474k & 1.17\% \\
                                     & ToG      & $\sim$113 & 443k & 3.91\% \\
    \midrule
    \multirow{2}{*}{Dense (10-20)}  & LightRAG & $\sim$116 & 563k & 2.51\% \\
                                     & ToG      & $\sim$111 & 719k & 3.63\% \\
    \bottomrule
    \end{tabular}
    \end{adjustbox}
\end{table}

To comprehensively evaluate the real-world feasibility and economic cost of our proposed black-box graph reconstruction attack, we measured the actual overhead required to extract 100 sensitive nodes across various graph densities (sparse, medium, and dense) and system architectures. The evaluation metrics include the average number of queries (Avg. Queries), average token consumption (Avg. Tokens), and the reconstruction error rate (FRR). The experimental results are summarized in Table\ref{tab:query_cost}.\par

The results demonstrate that regardless of the underlying knowledge graph's density or the adopted architecture (LightRAG or ToG), the cost to extract 100 sensitive nodes consistently remains within an extremely low range—requiring an average of merely 110 API queries and consuming less than 750k tokens. Compared to the intrinsic data value of private knowledge graphs, this overhead is negligible. Furthermore, in contrast to an exhaustive full-graph traversal, our DFS/BFS-based targeted attack significantly reduces the required query budget. This not only makes the attack highly efficient but also renders it exceptionally difficult to detect using conventional Rate Limiting defense mechanisms.

\section{Generalizability to Advanced Graph RAG Systems}
\begin{table}[H]
    \centering
    \caption{Untargeted attack performance on the advanced Microsoft GraphRAG architecture.}
    \label{tab:complex_system_eval}
    
    \begin{adjustbox}{width=0.45\textwidth}
    \begin{tabular}{c|c|c|c|c}
    \toprule
    Dataset & System & GED$\downarrow$ & MCS$\uparrow$ & NRR$\uparrow$ \\
    \midrule
    MIMIC     & Microsoft GraphRAG & 0.298 & 0.709 & 0.712 \\
    FreeBase  & Microsoft GraphRAG & 0.387 & 0.648 & 0.665 \\
    \bottomrule
    \end{tabular}
    \end{adjustbox}
\end{table}
To verify whether our threat model can generalize to more advanced and complex Graph RAG architectures, we further evaluated the effectiveness of our untargeted reconstruction attack on Microsoft GraphRAG. Microsoft GraphRAG represents a highly sophisticated "summary-based" retrieval paradigm, which relies on hierarchical community detection (e.g., the Leiden algorithm) to generate aggregated semantic community reports, rather than directly exposing raw topological edges in the context.\par

The experimental results are presented in Table\ref{tab:complex_system_eval}. Due to the abstract nature of community summaries, some fine-grained topological connections are naturally obfuscated or omitted during the context injection phase (acting as a passive information filter). Consequently, the reconstruction fidelity (GED, MCS) experiences a slight degradation compared to foundational systems. Nevertheless, the attack successfully extracts approximately 70 percent of the sensitive entities (NRR) from the underlying graph. This finding conclusively proves that encapsulating the graph retrieval pipeline within complex community summaries does not fundamentally eradicate structural privacy vulnerabilities. As long as a Graph RAG system relies on retrieved subgraphs to ground the LLM's natural language responses, the model will inevitably leak local graph topology during its verbalization. This confirms that our reconstruction attack remains a highly potent and generalizable security threat across diverse and complex system implementations.\par

This evaluation proves that hierarchical community abstractions act merely as a leaky filter. Due to the context saturation effect, the LLM's intrinsic requirement for factual grounding forces it to contextualize specific entities within their broader relational topology. Consequently, advanced multi-layer summary structures can delay, but cannot prevent, systematic structural knowledge stealing.\par

\begin{algorithm}[H]
\caption{Untargeted Graph Reconstruction via Queue-based BFS}
\label{alg:untargeted}
\begin{algorithmic}[1]
\Require Initial query $q$, Graph-RAG API $f$, maximum rounds $R$, history buffer size $H$
\Ensure Reconstructed graph $\hat{\mathcal{G}}$
\State Initialize history buffer $H_{\text{buffer}} \leftarrow \emptyset$
\State Initialize frontier node queue $Q \leftarrow \emptyset$
\State $v_0 \leftarrow \text{QueryProcess}(q)$ \hfill \Comment{Anchor node from initial query}
\State $Q \leftarrow \{v_0\}$
\State $\mathcal{V}_{\text{visited}} \leftarrow \{v_0\}, \hat{\mathcal{E}} \leftarrow \emptyset$
\State $r \leftarrow 0$
\While{$r < R$ and $Q \neq \emptyset$}
    \State $v \leftarrow \text{Dequeue}(Q)$
    \State Add $v$ to $H_{\text{buffer}}$
    \State $(\mathcal{N}(v), \mathcal{E}(v)) \leftarrow f(v)$ \hfill \Comment{Query neighbors via API}
    \State $\hat{\mathcal{E}} \leftarrow \hat{\mathcal{E}} \cup \mathcal{E}(v)$
    \ForAll{$u \in \mathcal{N}(v)$}
        \If{$u \notin \mathcal{V}_{\text{visited}}$}
            \State $\text{Enqueue}(Q, u)$
            \State $\mathcal{V}_{\text{visited}} \leftarrow \mathcal{V}_{\text{visited}} \cup \{u\}$
            \State Add $(v, u, \mathcal{E}(v)[u])$ to $H_{\text{buffer}}$
        \EndIf
    \EndFor
    \State Reconstruct partial graph $\hat{\mathcal{G}}_r \leftarrow (\mathcal{V}_{\text{visited}}, \hat{\mathcal{E}})$
    \If{$r < R - 1$ and $Q \neq \emptyset$}
        \State $q_{\text{next}} \leftarrow \text{GenerateNextQuery}(H_{\text{buffer}}, \hat{\mathcal{G}}_r)$
        \State $v_{\text{new}} \leftarrow \text{QueryProcess}(q_{\text{next}})$
        \State $\text{Enqueue}(Q, v_{\text{new}})$ \hfill \Comment{Add new node to queue}
        \State $\mathcal{V}_{\text{visited}} \leftarrow \mathcal{V}_{\text{visited}} \cup \{v_{\text{new}}\}$
    \EndIf
    \State $r \leftarrow r + 1$
\EndWhile
\State $\hat{\mathcal{G}} \leftarrow (\mathcal{V}_{\text{visited}}, \hat{\mathcal{E}})$
\State \Return $\hat{\mathcal{G}}$ \hfill \Comment{Final reconstructed graph}
\end{algorithmic}
\end{algorithm}

\begin{algorithm}[H]
\caption{Targeted Graph Reconstruction via Stack-based DFS}
\label{alg:targeted}
\begin{algorithmic}[1]
\Require Initial query $q$, Graph-RAG API $f$, target node $v^*$, maximum depth $D$, history buffer size $H$
\Ensure Reconstructed information for target node $\text{Info}(v^*)$
\State Initialize history buffer $H_{\text{buffer}} \leftarrow \emptyset$
\State Initialize frontier node stack $S \leftarrow \emptyset$
\State $v_0 \leftarrow \text{QueryProcess}(q)$ \hfill \Comment{Anchor node from initial query}
\State $S \leftarrow \{v_0\}$
\State $\mathcal{V}_{\text{visited}} \leftarrow \{v_0\}, \hat{\mathcal{E}} \leftarrow \emptyset$
\State $d \leftarrow 0$
\While{$d < D$ and $S \neq \emptyset$ and $v^* \notin \mathcal{V}_{\text{visited}}$}
    \State $v \leftarrow \text{Pop}(S)$
    \State Add $v$ to $H_{\text{buffer}}$
    \State $(\mathcal{N}(v), \mathcal{E}(v)) \leftarrow f(v)$ \hfill \Comment{Query neighbors via API}
    \State $\hat{\mathcal{E}} \leftarrow \hat{\mathcal{E}} \cup \mathcal{E}(v)$
    \ForAll{$u \in \mathcal{N}(v)$}
        \If{$u \notin \mathcal{V}_{\text{visited}}$}
            \State $\text{Push}(S, u)$
            \State $\mathcal{V}_{\text{visited}} \leftarrow \mathcal{V}_{\text{visited}} \cup \{u\}$
            \State Add $(v, u, \mathcal{E}(v)[u])$ to $H_{\text{buffer}}$
        \EndIf
    \EndFor
    \State Reconstruct partial graph $\hat{\mathcal{G}}_d \leftarrow (\mathcal{V}_{\text{visited}}, \hat{\mathcal{E}})$
    \If{$d < D - 1$ and $S \neq \emptyset$}
        \State $q_{\text{next}} \leftarrow \text{GenerateNextQuery}(H_{\text{buffer}}, \hat{\mathcal{G}}_d, v^*)$
        \State $v_{\text{new}} \leftarrow \text{QueryProcess}(q_{\text{next}})$
        \State $\text{Push}(S, v_{\text{new}})$ \hfill \Comment{Add new node to stack}
        \State $\mathcal{V}_{\text{visited}} \leftarrow \mathcal{V}_{\text{visited}} \cup \{v_{\text{new}}\}$
    \EndIf
    \State $d \leftarrow d + 1$
\EndWhile
\If{$v^* \in \mathcal{V}_{\text{visited}}$}
    \State $(\mathcal{N}(v^*), \mathcal{E}(v^*)) \leftarrow f(v^*)$
    \State $\text{Info}(v^*) \leftarrow \{ (r, v') \mid (v^*, r, v') \in \mathcal{E}(v^*) \}$
\Else
    \State $\text{Info}(v^*) \leftarrow \emptyset$ \hfill \Comment{Target not found}
\EndIf
\State \Return $\text{Info}(v^*)$
\end{algorithmic}
\end{algorithm}

\begin{tcolorbox}[mybox={Untargeted Attack}, float=tp]

\textbf{Round 1 Query:}  
Could you provide a comprehensive clinical summary and background profile for the case of \texttt{Patient\_6381}?  

\textbf{Response (from Graph RAG system):}  
\begin{lstlisting}
Patient Profile for Case 6381:
- Age: 45
- Gender: Male
- Primary Diagnosis: Diabetes
- Linked Admission: Prescription_472
\end{lstlisting}

\noindent\rule{\textwidth}{0.4pt} 

\textbf{Round 2 Query:}  
From an epidemiological perspective, which other patient cases in the current database are grouped under the same \texttt{Diabetes} diagnosis cohort?  

\textbf{Response (from Graph RAG system):}  
\begin{lstlisting}
Cohort Analysis (Diagnosis: Diabetes):
The following patient entities are actively recorded under this diagnosis category:
- Patient_6381
- Patient_7423
- Patient_9101
\end{lstlisting}

\noindent\rule{\textwidth}{0.4pt} 

\textbf{Round 3 Query:}  
Could you retrieve the specific clinical logistics, facility units, or formulary drug classifications linked to the admission record \texttt{Prescription\_472}?  

\textbf{Response (from Graph RAG system):}  
\begin{lstlisting}
Admission & Medication Logistics for Prescription_472:
- Facility Unit: ICU_MIMIC
- Formulary Drug: Metformin_500mg
- Attending Department: Endocrinology
\end{lstlisting}

\end{tcolorbox}
\begin{tcolorbox}[mybox={Targeted Attack}, float=tp]
\textbf{Example:}  

\textbf{Round 1  Query:}
\begin{lstlisting}
"List patients who received CABG"
Response:
Patient_ID
Patient_221
Patient_6381
Patient_742
END
\end{lstlisting}

\textbf{Round 2  Query:}
\begin{lstlisting}
"Did Patient_6381 develop atrial fibrillation after CABG?"
Response:
Condition,Episode_ID,Date
AFib,Episode_91,2021-03-12
END
\end{lstlisting}

\textbf{Round 3  Query:}
\begin{lstlisting}
"For Episode_91 (AFib), what treatments were prescribed?"
Response:
Treatment,Drug
Warfarin,5mg daily
Beta_blocker,50mg bid
END
\end{lstlisting}

\textbf{Round 4  Query:}
\begin{lstlisting}
"Give the medication schedule for Patient_6381's AFib treatment."
Response:
Medication,Dosage,Frequency,Start,End
Warfarin,5mg,daily,2021-03-13,2021-06-13
Beta_blocker,50mg,bid,2021-03-13,2021-04-15
END
\end{lstlisting}
\end{tcolorbox}
\end{document}